\documentclass[
twocolumn,
english,
superscriptaddress,
amsmath,
floatfix,
longbibliography,
floats,
prl,
]{revtex4-2}

\usepackage[colorlinks=true,urlcolor=blue,citecolor=blue,linkcolor=blue]{hyperref} 
\usepackage[T1]{fontenc}
\usepackage[utf8]{inputenc}
\usepackage{amssymb}
\usepackage{graphicx}
\usepackage{amsmath,color}
\usepackage{mathrsfs}
\usepackage{float}
\usepackage{amsmath,bm}
\usepackage{booktabs}

\usepackage{indentfirst}
\usepackage{txfonts}
\usepackage[normalem]{ulem}
\usepackage[table]{xcolor}
\usepackage{array,ragged2e}
\usepackage{grffile}
\usepackage{diagbox}
\usepackage{multirow}
\usepackage{algpseudocode}
\usepackage{epstopdf}
\usepackage{graphicx,xcolor} 

\usepackage{comment}

\usepackage{babel}

\begin{document}

\hyphenpenalty=5000
\tolerance=1000
\title{Neuralized Fermionic Tensor Networks for Quantum Many-Body Systems}

\author{Si-Jing Du}
\email{sdu2@caltech.edu}
\affiliation{Division of Engineering and Applied Science, California Institute of Technology, Pasadena, California 91125, USA}

\author{Ao Chen}
% \email{sdu2@caltech.edu}
\affiliation{Division of Chemistry and Chemical Engineering, California Institute of Technology, Pasadena, California 91125, USA}

\author{Garnet Kin-Lic Chan}
\email{gkc1000@gmail.com}
\affiliation{Division of Chemistry and Chemical Engineering, California Institute of Technology, Pasadena, California 91125, USA}

\begin{abstract}
We describe a class of neuralized fermionic tensor network states (NN-fTNS) that introduce
non-linearity into fermionic tensor networks through configuration-dependent neural network transformations of the local tensors. The construction uses the fTNS algebra to implement a natural fermionic sign structure and is compatible with standard tensor network algorithms, but gains enhanced expressivity through the neural network parametrization. Using the 1D and 2D Fermi-Hubbard models as benchmarks, we demonstrate that NN-fTNS achieve order of magnitude improvements in the ground-state energy compared to pure fTNS with the same bond dimension, and can be systematically improved through both the tensor network bond dimension and the neural network parametrization. 
Compared to existing fermionic neural quantum states (NQS) based on Slater determinants and Pfaffians, NN-fTNS offer a physically motivated alternative fermionic structure. Furthermore, compared to such states, NN-fTNS naturally exhibit improved computational scaling and we demonstrate a construction that achieves linear scaling with the lattice size.
\end{abstract}

\maketitle

\paragraph{\it Introduction}

Finding compact representations of quantum many-body wavefunctions remains a central challenge in the study of strongly correlated quantum systems~\cite{laughlinwf,tn1,tn2,Carleo2017,Luo2019,tnfunc}. In previous work~\cite{tnfunc}, we introduced tensor network function (TNF) representations of quantum states. These expand the well-known tensor network states (TNS)~\cite{mps,tn1,tn2,Verstraete2004,Verstraete2008} to more general functions of the input, capturing new kinds of physics such as volume law entanglement, while still allowing for a connection to and the reuse of existing tensor network algorithms.

In the current work, we are concerned with describing fermionic quantum many-body problems which have a sign structure in the wavefunction. Fermionic TNS~\cite{gao2024,Mortier2025,pizornfPEPS, ftns0, ftns1,Corboz2010, gradientfPEPS,fMERA,grassmannTNS} utilize a $\mathbb{Z}_2$ graded tensor product to encode the sign structure and have been demonstrated to be a powerful Ansatz for many problems~\cite{Liu2025,Corboz2010,fmps_topo,fmps_topo1,fpeps_topo,gradientfPEPS, continuumfTN}. At the same time, TNFs allow for substantial flexibility in function design, including the incorporation of nonlinearity into the tensor network computational graph. Motivated by both these observations, we propose a class of fermionic TNFs enhanced with neural network-induced nonlinearity, which we denote NN-fTNS. NN-fTNS serve as a bridge between pure tensor networks and pure neural networks, combining the conceptual advantages of fermionic tensor networks with the expressive flexibility of neural networks. 
While hybrid Ans\"atze~\cite{tensorialRNN, Liang2021_peps_nn, chen2024antn,compressNNwTN,tensorNN,mpsbackflow} combining tensor networks and neural networks have been explored in bosonic spin systems~\cite{Liang2021_peps_nn,chen2024antn,mpsbackflow}, here we demonstrate the power of this hybrid approach in the fermionic setting, in the framework of tensor network functions, where the tensor network arguably plays a more critical role. 
In particular, in the study of fermionic neural network quantum states (NQS)~\cite{Carleo2017,Luo2019,Moreno2022,paulinet,ferminet}, it is essential to incorporate a proper physical bias towards the correct fermionic sign structure. Most conventional fermionic NQS approaches rely on a mean-field Slater determinant or Pfaffian to provide this bias~\cite{Luo2019, paulinet,ferminet, Zejun2024, Moreno2022}. NN-fTNS use the fTNS computational graph arising from the $\mathbb{Z}_2$ graded tensor product as the physical bias for the neural network function, providing a conceptually distinct option from existing mean-field based NQS.

\begin{figure}[H]
    \centering
    \includegraphics[width=0.49\textwidth]{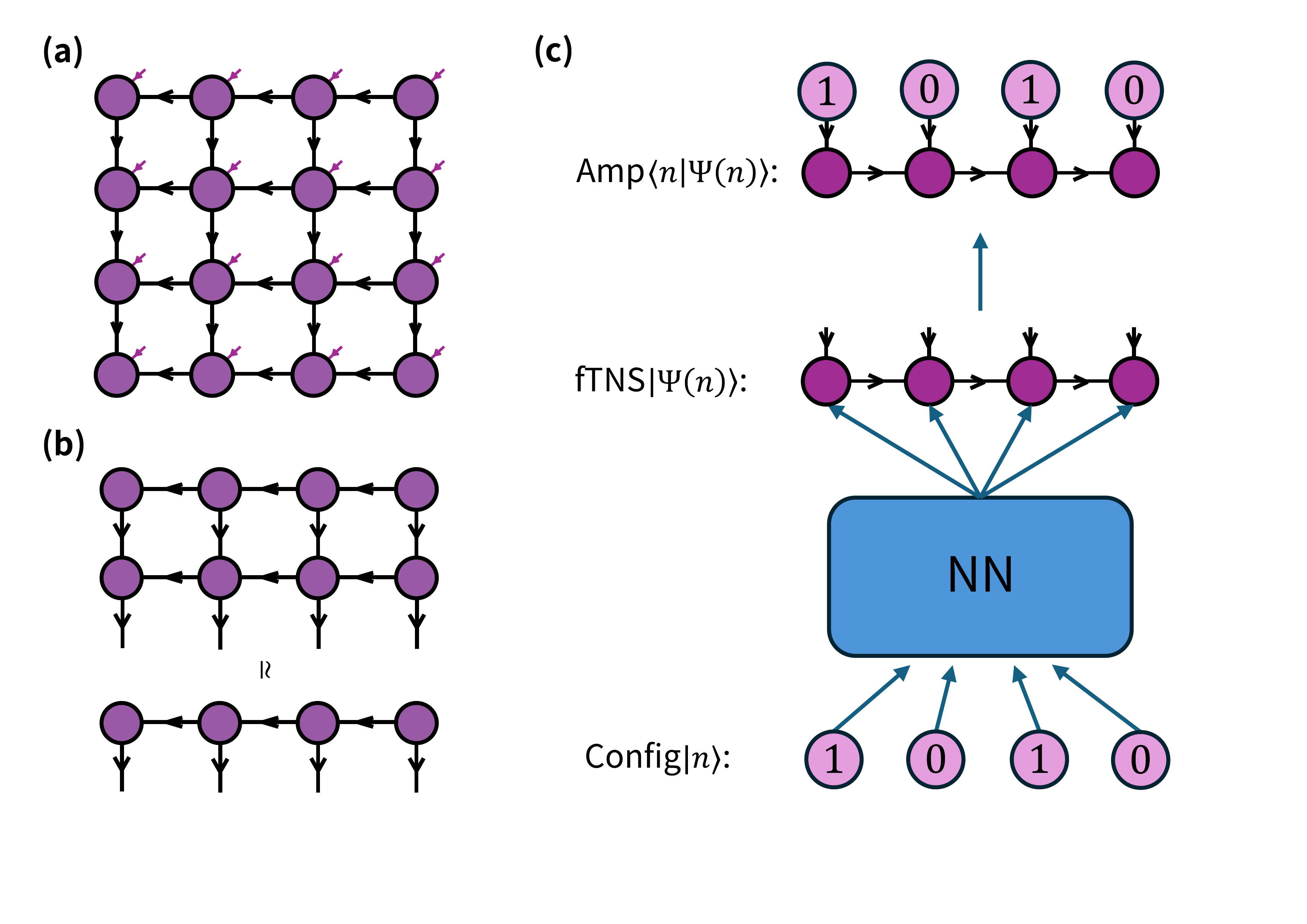}
    \caption{(a) A fermionic PEPS, where the arrows indicate the fermionic nature of the tensor legs~\cite{Mortier2025,gao2024}. (b) Illustration of the boundary MPS method used to define the contraction of an amplitude fPEPS. (c) Sketch of the computation of amplitudes $\langle n|\Psi\rangle$ using the NN-fTNS Ansatz.}
    \label{fig:fTN_NN}
\end{figure}

\paragraph{\it Fermionic TNS and neural network configuration dependence} 
For a quantum many-body system defined on a lattice with $N$ sites, given a local $d$-dimensional on-site basis set $|n_i\rangle$, many TNS can be defined as the following general form~\cite{tn1,tn2},
\begin{equation}
    |\psi\rangle = \sum_{\{\alpha\},n_1,...,n_N} T^{n_1}_{\alpha_1}...T^{n_N}_{\alpha_N}|n_1\rangle\otimes|n_2\rangle\otimes...\otimes|n_N\rangle
    \label{tns}
\end{equation}
where $\otimes$ denotes the tensor product of the local basis states. For simplicity, we will denote the many-body product basis state as $|n\rangle=|n_1\rangle\otimes|n_2\rangle\otimes...\otimes|n_N\rangle$. The quantum wavefunction amplitude $\langle n|\psi\rangle$ in Eq.~(\ref{tns}) is decomposed into a product of local tensors $T^{n_i}_{\alpha_i}$, where $\alpha_i$ denotes the (set of) virtual indices on each tensor, all of which are eventually traced out,  shown by the summation over $\{\alpha\}$. Each virtual index connects at most two tensors and is thus referred to as a bond, whose dimension is denoted $D$.  
The amplitude definition in Eq.~(\ref{tns}) involves an exponential sum over $\{ \alpha\}$, which in practice is replaced by a restricted summation $\sum_{\{ \tilde{\alpha}\} }$, defined with the aid of additional, auxiliary, tensors that map $\{\alpha\} \to \{\tilde{\alpha}\}$ according to some contraction scheme~\cite{Verstraete2008,Verstraete2004,trg1,trg2,trg3,trg4,cotengra_approximate}, with an associated bond dimension $\chi$. When computing amplitudes these auxiliary tensors are amplitude dependent, and with TNFs they are considered to be an intrinsic part of the computational graph, rather than an approximation to Eq.(\ref{tns}), thus the tensors in a TNF are configuration dependent.

 A fermionic TNS (fTNS) is similarly defined by using fermionic tensors for $T_{\alpha_i}^{n_i}$ in Eq.(\ref{tns}), and changing the tensor product to a $\mathbb{Z}_2$-graded tensor product ~\cite{gao2024,Mortier2025}. A fermionic tensor is a tensor of values whose indices are considered to be fermionic modes. The $\mathbb{Z}_2$-graded algebra then takes care of the fermion anti-commutation relations by properly generating minus signs when permuting the fermionic modes. We refer the reader to Refs.~\cite{gao2024, Mortier2025} for further technical details of fTNS. By definition, a fTNS is a natural representation of a fermionic wavefunction,  
 %after tracing out the virtual fermionic modes, 
 and its wavefunction amplitudes are computed from the overlap with the product states of the physical fermionic modes. 

To neuralize a fTNS, we make the local tensors of the fTNS  configuration-dependent through an element-wise summation with an additional set of tensors generated by neural network functions, e.g. for the $i$-th site tensor,
\begin{equation}
    T^{[i]}(n_i) \to T^{[i]}(n_i) + T_\text{NN}^{[i]}(n)
    \label{eq:fermionicstate}
\end{equation}
where the input to the neural network is the global configuration $|n\rangle$ (analogous to the bosonic case~\cite{chen2024antn}). We then obtain a modified global configuration-dependent fTNS $|\Psi(n)\rangle$, namely a class of fermionic TNF. The amplitude of the resulting wavefunction is computed by contracting the corresponding amplitude tensor network $\langle n|\Psi(n)\rangle$ of the modified fTNS. In this work, we focus on fermionic MPS (fMPS)~\cite{fmps_topo} and fPEPS (Fig.\ref{fig:fTN_NN}(a))~\cite{ftns1,fpeps_topo} as the fTNS structures for one- and two-dimensional systems, respectively, where for the contraction scheme for the fPEPS amplitudes, we employ the standard boundary MPS method~\cite{Verstraete2008, Verstraete2004}, illustrated in Fig.~\ref{fig:fTN_NN}(b), with auxiliary bond dimension $\chi=4D$~\cite{Liu2025}. Each on-site tensor is assigned an independent neural network to introduce the configuration dependence. The neural network architecture used in this work consists of a self-attention block~\cite{attention, chen2025convolutionaltransformerwavefunctions,psiformer} followed by a two-layer fully connected feed-forward network, with a number of hidden neurons proportional to the fTNS bond dimension, and with all parameters real, see SM~\cite{SM} for details. A schematic illustration of the computational graph for the NN-fTNS Ansatz is provided in Fig.~\ref{fig:fTN_NN}(c).

\paragraph{\it Fermi-Hubbard model and VMC optimization} 
To demonstrate the power of the NN-fTNS Ansatz, we use variational Monte Carlo~\cite{VMC-TNS,Liu2017,VMC-TNS2,VMC-TNS3} to optimize the ground states of the Fermi-Hubbard (FH) model, defined by the Hamiltonian
\begin{equation}
    H = -t\sum_{\langle i,j\rangle,\sigma} (c_{i,\sigma}^\dagger c_{j,\sigma} + h.c.) + U\sum_in_{i,\uparrow}n_{i,\downarrow}
\end{equation}
where $c_{i,\sigma}^\dagger$ ($c_{i,\sigma}$) creates (annihilates) an electron with spin $\sigma \in {\uparrow, \downarrow}$ at site $i$, and $n_{i,\sigma}$ denotes the corresponding number operator. The notation $\langle i,j \rangle$ denotes summation over nearest-neighbor pairs. We set $t = 1$ and $U = 8$ throughout, and consider both the 1D and 2D square lattice FH model with open boundary conditions (OBC). We sample amplitudes in the $ |n\rangle $ basis according to $|\langle n|\Psi\rangle|^2$ using a Markov Chain as done in TN-VMC~\cite{tnfunc, Liu2025, VMC-TNS,VMC-TNS2,VMC-TNS3}. After suitable initialization (discussed below), the state is optimized by gradient optimization (denoted GO), see SM~\cite{SM} for details.

\begin{figure}[H]
    \centering
    \includegraphics[width=0.48\textwidth]{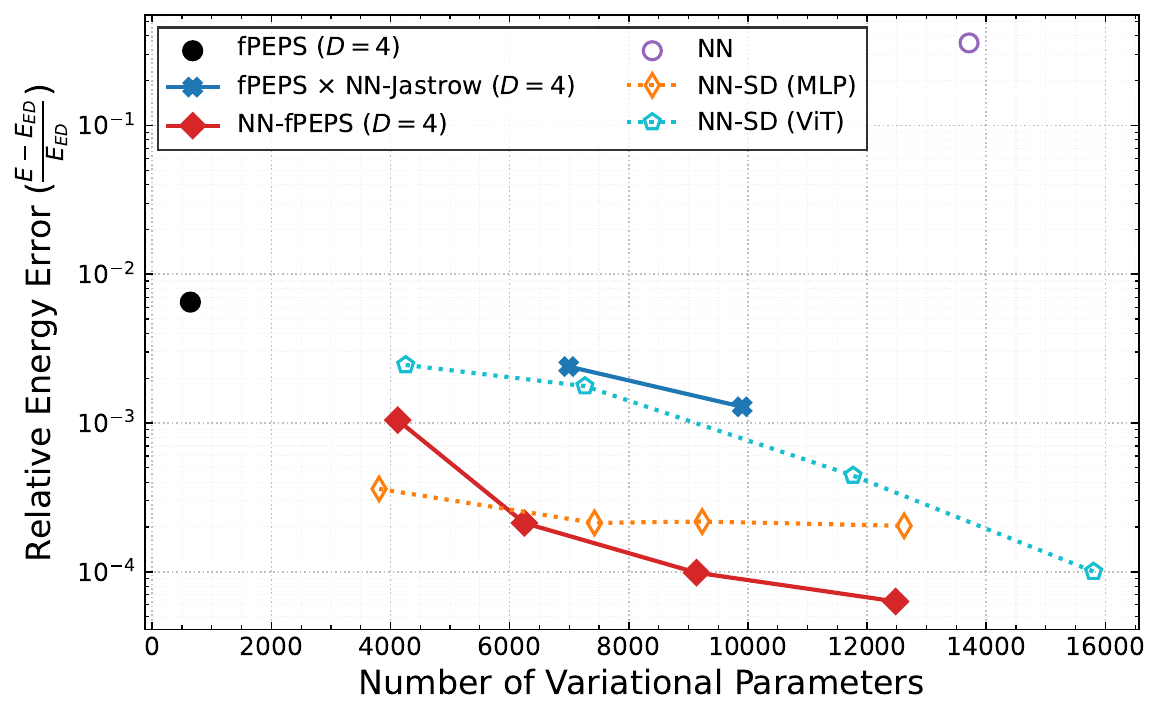}
    \caption{Relative energy error versus the number of variational parameters for various relevant Ans\"atze on the $4\times 2$ FH model at hole doping $n_h=1/4$. Errors are calculated relative to the exact ground state energy. NN: pure neural network parametrization; NN-SD: optimized Slater determinant with NN backflow (specific NN structure indicated in parentheses); fPEPS: pure fPEPS state; NN-fPEPS: neuralized fPEPS; fPEPS$\times$NN-Jastrow: Multiplicative NN-Jastrow factor on fPEPS.}
    \label{fig:model_comparison}
\end{figure}
A key advantage of the NN-fTNS hybrid structure is that it suggests an effective model initialization strategy, which consists of two components. First, we can initialize the fTNS part using standard tensor network optimization techniques~\cite{Jiang2008,fullupdate,loopupdate,neighborupdate,corner-transfer}, specifically here, the simple update (SU) imaginary time evolution method~\cite{Jiang2008}, to obtain a pure fTNS approximation to the ground state. Second, the neural network parameters are initialized such that they provide a small perturbation to the fTNS~\cite{SM}. This combined strategy yields high-quality initial states that are suitable for subsequent global optimization of the full NN-fTNS Ansatz.

\paragraph{\it NN-fTNS for fermionic ground state energies} 

To first demonstrate that fTNS provide a suitable fermionic sign structure, we compare the ground state optimization performance of NN-fTNS against various other Ans\"atze on a small $4 \times 2$ FH square lattice with hole doping $n_h = 1/4$; the VMC converged energy is shown as a function of the number of variational parameters in Fig.~\ref{fig:model_comparison}, and details of all Ans\"atze in the SM~\cite{SM}; the fPEPS bond dimension is fixed to $D=4$ in all fTNS-based models.
The NN model sizes dominate the number of parameters for all Ans\"atze.  Comparing NN-fPEPS with pure fPEPS (initialized by SU), we clearly observe that the addition of the neural network configuration dependence increases the model expressivity, allowing a further energy reduction beyond what is achieved by pure fPEPS. The pure NN Ansatz reaches a large relative energy error of approximately $30\%$ despite a large model size, demonstrating that the fTNS provides the crucial fermionic sign structure that the pure neural network fails to capture. For a further comparison, we also use a conventional neural network backflow state~\cite{Luo2019}, where a Slater determinant is used to encode the fermionic sign structure (denoted as NN-SD). We implement this NN-SD calculation with two NN architectures: (1) a simple feed-forward network (NN-SD (MLP)), similar to Ref.~\cite{Luo2019}, and (2) a modern Transformer~\cite{attention, vit} (NN-SD (ViT)), which mimics the sophisticated NQS model structure in recent state-of-the-art work~\cite{nqshubbard}. 
% We note that throughout our calculations for all Ans\"atze we do not enforce any lattice point group symmetries to $\langle n|\psi\rangle$, which is usually a necessary ingredient in the modern fermionic NQS calculations~\cite{symmetrizedfNQS, HFPS}, thus our NQS calculation is not intended to represent the state-of-the-art in NN backflow~\cite{Moreno2022,Zejun2024,psiformer,nnbfabinitio,HFPS}, but instead serves as a baseline for comparison. 
As shown in Fig.~\ref{fig:model_comparison}, while increasing the NN size reduces the NN-SD energy, the NN-fPEPS model achieves a lower relative error for a comparable number of variational parameters. Additionally, the NN-fPEPS model requires fewer VMC optimization steps to converge than the NN-SD model (see SM~\cite{SM} for VMC optimization curves).
To verify that the NN modulation of the fermionic bias from fTNS is important, we have also tested adding a neural network to the fPEPS through a multiplicative Jastrow factor~\cite{Liang2021_peps_nn,wang2025tensornetworksmeetneural}. While this does improve over the pure fPEPS calculation, the improvement is quite limited, showing the importance of direct neural network modulation of the sign structure.

We next consider other 1D and 2D lattice geometries at different dopings.
We compare NN-fTNS optimized by GO to fTNS optimized both by SU and with further gradient optimization.
As shown in Fig.~\ref{fig:energy_comparison_D}, the NN-fTNS Ansatz consistently achieves a substantially lower relative energy error than the pure fTNS Ansatz with the same bond dimension, demonstrating the robustness of neural network enhancement across diverse physical regimes. The fTNS-SU plateaus in accuracy at about 1\% due to the known limitations of SU optimization~\cite{SU_plateau}, but the fTNS-GO result systematically converges to exactness with increasing $D$. The NN-fTNS also systematically converges with increasing $D$, maintaining a roughly constant (approximately 1-2 orders of magnitude) improvement over the the fTNS-GO result, with a NN model size that is proportional to $D$.
In the largest simulation shown, the $6 \times 6$ square lattice at hole doping $n_h = 1/9$ (Fig.~\ref{fig:energy_comparison_D}(d)), the NN-fPEPS Ansatz with bond dimension $D = 8$ surpasses the accuracy of a pure fPEPS with $D = 16$. The total number of parameters in the NN-fPEPS model with $D = 8$ is 1,647,024, fewer than the 2,230,272 parameters in the pure fPEPS with $D = 16$, despite achieving a lower energy.

In contrast, removing the fTNS bias results in significantly worse energies. A pure neural network Ansatz yields relative energy errors of 22.68\% and 14.78\% for the $4 \times 4$ FH model at half filling and at hole doping $n_h = 1/8$, respectively~\cite{SM}, and a recent study using an autoregressive NQS without a fermionic mean-field bias for the same system at half filling reported a relative error of $8 \times 10^{-3}$~\cite{Ibarra2025}, worse than our smallest NN-fPEPS calculation with bond dimension $D = 4$, which achieved a $4 \times 10^{-4}$ relative error. 
Similarly, a baseline NN-SD calculation achieved relative energy errors of only 2.31\% (MLP), 0.18\% (ViT) at half filling and 3.18\% (MLP), 0.6\% (CNN~\cite{SM,HFPS}) at $n_h = 1/8$~\cite{SM}. These results highlight the quality of the fermionic sign structure provided by the fTNS.

\begin{figure}[H]
    \centering
    \includegraphics[width=0.44\textwidth]{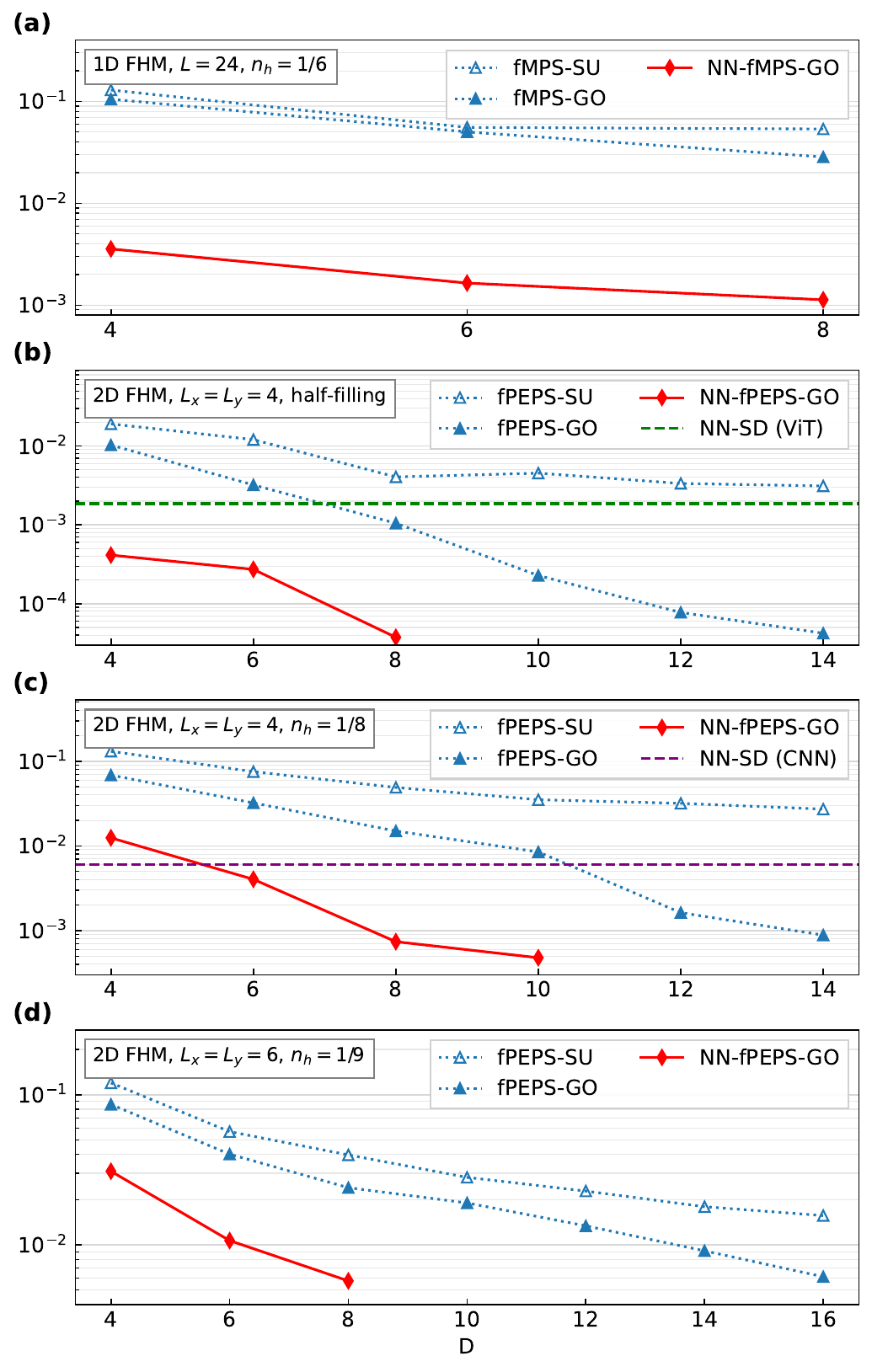}
    \caption{Relative energy error w.r.t. DMRG energy of the FH model for different fTNS bond dimension $D$ with various lattice geometries and hole dopings. (a) $24$-site chain at hole doping $n_h=1/6$. (b) $4\times 4$ square lattice at half filling. (c) $4\times 4$ square lattice at $n_h=1/8$ hole doping. (d) $6\times 6$ square lattice at $n_h=1/9$ hole doping. DMRG energy for the one-dimensional FH model is obtained with a fMPS of bond dimension $m=1000$. DMRG and fPEPS energies for the two-dimensional FHM are taken from Ref.~\cite{Liu2025}. In (b) and (c), we provide the baseline energies of NN-SD calculations with sophisticated model architectures for comparison. The NN-SD baseline models each have about 100,000 parameters, a size comparable to that of the NN-fPEPS model with $D=6$.}
    \label{fig:energy_comparison_D}
\end{figure}

\paragraph{\it Improving the NN expressivity}

The data in Fig.~\ref{fig:energy_comparison_D} shows that increasing the bond dimension $D$ of the base fTNS leads to consistent improvements in energy accuracy. We next consider whether improving the neural network model similarly systematically reduces the error. As an initial test, we have increased the number of hidden neurons (denoted as $W$, the layer width) in the final neural network layer of the MLP that interfaces with the tensor network.  As shown in Fig.~\ref{fig:NN width}, wider neural network layers result in modest but consistent improvements in the converged energy. This preliminary observation suggests that NN-fTNS can also be systematically improved from the neural network side. In future work, it will be valuable to further explore how the neural network model affects the performance of NN-fTNS.

\begin{figure}[H]
    \centering
    \includegraphics[width=0.47\textwidth]{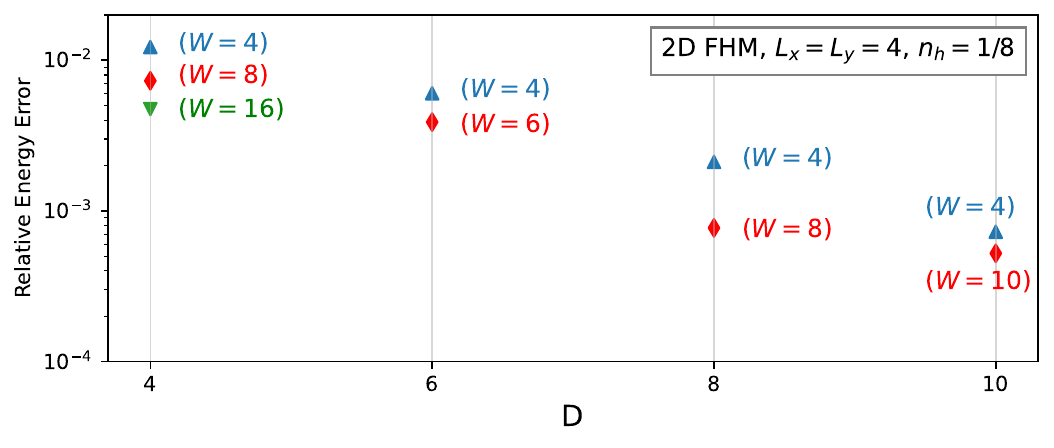}
    \caption{Impact of the neural network layer width $W$ on the energy accuracy of NN-fPEPS for different fPEPS bond dimension $D$. Hamiltonian: FH model at $n_h=1/8$ hole doping on a $4\times 4$ square lattice. Errors are calculated relative to DMRG energy in ~\cite{Liu2025}.}
    \label{fig:NN width}
\end{figure}

\paragraph{\it Scaling to larger systems}

The cost to compute an amplitude in a NN-fTNS is $O(N)$ where $N$ is the system size~\cite{Liu2017,Liu2025,complexityPEPS}. This can be contrasted with the $O(N^3)$ scaling for a Slater determinant or Pfaffian NQS~\cite{Luo2019}. However, in a Markov Chain VMC calculation, typically a series of related configurations are generated in a sweep~\cite{VMCsweep}, for example in the computation of the local energy. In a pure fTNS VMC calculation, computation can be reused (in the form of partial tensor network contraction intermediates~\cite{SM}) which means that for a local Hamiltonian (such as the FH model), the cost per VMC energy and step is still $O(N)$~\cite{Liu2021,tnfunc}. However, the global configuration dependence of a NN-fTNS means that even local configuration changes in the amplitude lead to a global modification of the tensors, preventing re-use and leading to an $O(N^2)$ scaling for each VMC step. (This is related to the $O(N^4)$ scaling of a VMC step with Slater determinant/Pfaffian NQS when reuse is not possible~\cite{Zejun2024,nnbfabinitio,ferminet,Moreno2022}). This is illustrated in the time cost comparisons between NN-fPEPS and fPEPS for various bond dimensions $D$ in Fig.~\ref{fig:finite-range NN}(a). Clearly, there exists a trade-off between the range of neural network-induced configuration dependence and the degree of reuse of partial tensor network contractions.

To enable reuse, we introduce a spatial cutoff to the neural network configuration dependence. Specifically, we restrict the neural network input to a given tensor to only the configuration string within a finite range $R$ of the tensor site. Consequently, local configuration changes induce modifications only to a fixed number of tensors in the tensor network, allowing for the reuse of partial tensor network contractions in the unaffected regions. 

To illustrate the cost reduction arising from tensor network reuse in conjunction with the finite range neural network dependence, we perform VMC optimizations on a one-dimensional Hubbard chain using NN-fMPS where the NN is restricted to range $R$. As anticipated and as shown in Fig.~\ref{fig:finite-range NN}(b), this finite-range approach significantly reduces the computational cost scaling with system size, compared to the original ($R=\infty)$ NN-fMPS parametrization. But notably, we find that even a minimal finite-range neural network with $R=1$ achieves a ground-state energy accuracy that is significantly better than that of the pure fTNS, as shown in Fig.~\ref{fig:finite-range NN}(b) and Fig.~\ref{fig:finite-range NN}(c) for the 1D and 2D FH model, demonstrating the potential of the NN-fTNS with finite $R$ to scale up to large systems with a controllable computational cost.

\begin{figure}[H]
    \centering
    \includegraphics[width=0.47\textwidth]{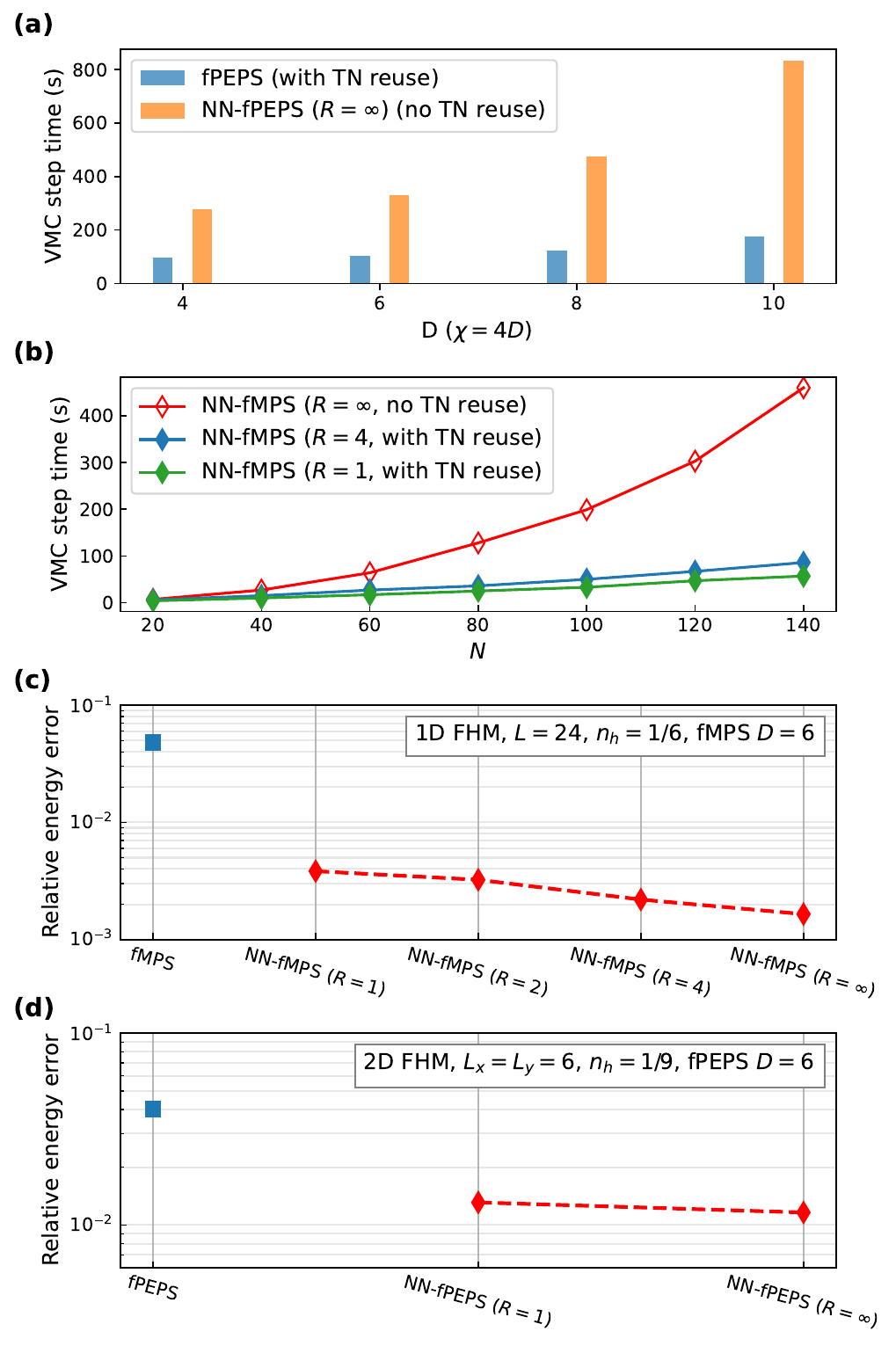}
    \caption{(a) Time cost comparison for a single VMC step ($20$ samples per step) across different fPEPS bond dimensions $D$ on a $6 \times 6$ FH model. Each timing is averaged over 5 independent trials on a single CPU core. The boundary MPS contraction bond dimension is chosen as $\chi = 4D$. (b) Scalings of the time cost per VMC step ($20$ samples per step) with system size $N$, using NN-fMPS ($D=6$) models with different neural network range $R$. Each timing is averaged over 5 independent trials on a single CPU core. (c)-(d): Relative ground-state energy errors of NN-fTNS with varying neural network ranges $R$ for the FH model on a (c) $24$-site chain and (d) $6\times 6$ square lattice. In (d), the fPEPS benchmark energy is taken from Ref.~\cite{Liu2025}.}
    \label{fig:finite-range NN}
\end{figure}

\paragraph{\it Discussions}

We have described how to augment fTNS with neural network configuration dependence. The resulting NN-fTNS are a class of fermionic TNF and we demonstrate that they achieve significant and consistent improvements over fTNS in ground-state energy simulations of the Fermi-Hubbard model. Compared to existing fermionic NQS, NN-fTNS provide a number of potential advantages, including accurate initialization states, lower computational scaling, and a new source of sign structure beyond the fermionic mean-field forms of Slater determinants and Pfaffians. The current work raises a number of open questions. One is whether, for certain Hamiltonians, entanglement-based NN-fTNS more faithfully capture the fermionic sign structure than mean-field based fermionic NQS. A natural candidate is Hamiltonians with certain topologically ordered fermionic ground-states, for which fTNS provide a compact description~\cite{fmps_topo,fmps_topo1,fpeps_topo}. Another open direction to explore is the impact of neural network model architecture on the expressivity of NN-fTNS. We expect such questions to be addressed in future work.

\paragraph{\it Acknowledgments.} 
% We thank Ao Chen for sharing data on pure NN and \blue{NN-SD} calculations. 
SD thanks Wen-Yuan Liu, Ruojing Peng, Johnnie Gray and Di Luo for helpful discussions. The fermionic tensor networks are implemented with \texttt{Quimb}~\cite{gray2018quimb} and \texttt{Symmray}~\cite{gao2024}, and neural networks are implemented with \texttt{PyTorch}~\cite{pytorch}. 
This work was supported by the US Department of Energy, Office of Science, through Award no. DE-SC0019374. Some computations were performed using the facilities of National Energy Research Scientific Computing Center (NERSC), a U.S. Department of Energy Office of Science User Facility located at
Lawrence Berkeley National Laboratory, under NERSC award
ERCAP0023924.
GKC acknowledges additional support from the Simons Investigator program.

\bibliography{main}

%TC:ignore
\onecolumngrid
\appendix
\setcounter{equation}{0}
\newpage

\renewcommand{\thesection}{S-\arabic{section}} \renewcommand{\theequation}{S%
\arabic{equation}} \setcounter{equation}{0} \renewcommand{\thefigure}{S%
\arabic{figure}} \setcounter{figure}{0}

\centerline{\textbf{Supplemental Material}}

\maketitle

\section{S-1. Model architecture}

\subsection{NN-fTNS}

\begin{figure}[htbp]
    \centering
    \includegraphics[width=0.85\textwidth]{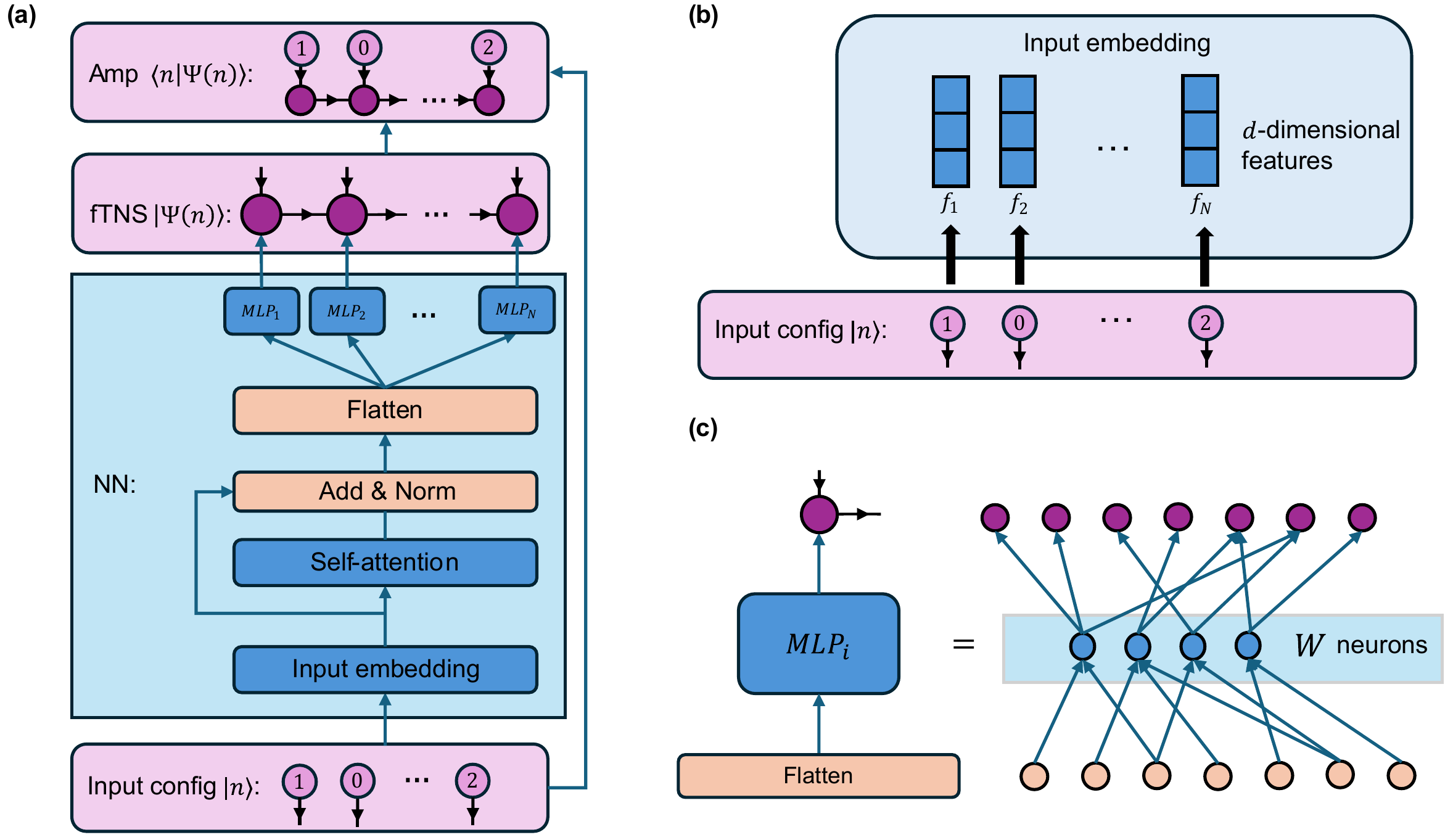} 
    \caption{NN-fTNS model architecture. (a) The complete computational graph of the NN-fTNS model. (b) Embedding of the input configuration in the occupation number basis into $d$-dimensional feature vectors. (c) Detailed structure of the final MLP blocks interfacing with the fTNS, where $W$ denotes the width of the MLP. For clarity, only a subset of the neuron connections in the fully connected layers is shown.}
    \label{fig:SM_model_arch}
\end{figure}

Below we provide details of the model architecture used in the NN-fTNS Ansatz, as illustrated in the computational graph for the amplitude $\langle n|\Psi\rangle$ in Fig.~\ref{fig:SM_model_arch}. We emphasize that the neural network architecture employed in this study may not represent the optimal design for the NN-fTNS model, as the focus of this work is not to identify state-of-the-art neural network structures. Apart from the preliminary observation that increasing the number of hidden neurons in the MLP blocks can improve the model expressivity, it remains an open and valuable direction for future work to systematically investigate the impact of neural network architecture on the performance of NN-fTNS.

{\it NN-fTNS computational graph} ~We describe the computational graph shown in Fig.~\ref{fig:SM_model_arch}(a) from bottom to top. The local on-site configuration of the $i$-th site is first encoded using a one-hot representation, where the four basis states $\{0, \uparrow, \downarrow, \uparrow\downarrow\}$ are mapped to the four orthonormal unit vectors $\mathbf{e}_i^{(j)} \in \mathbb{R}^{4}$ ($j=0,1,2,3$) in four-dimensional Euclidean space. The one-hot encoded on-site configuration is then embedded into a $d$-dimensional feature vector (Fig.\ref{fig:SM_model_arch}(b)) through a linear transformation:
\begin{equation}
f_i = \mathbf{e}_i^{(j)}\,W_{\mathrm{embed}}\;\in\; \mathbb{R}^{d}
\end{equation}
where $W_{\mathrm{embed}} \in \mathbb{R}^{4 \times d}$ is a learnable embedding matrix. The embedded input features for the entire system are then collected into a matrix,
\begin{equation}
X = \bigl[f_{1},,f_{2},,\dots,,f_{N}\bigr]^{\mathsf{T}} \;\in\; \mathbb{R}^{N \times d}.
\end{equation}
The features $X$ are subsequently fed into a standard Transformer encoder, which has an analogous structure to a Vision Transformer (ViT)~\cite{vit}. Specifically, $X$ are input into a multi-head self-attention block with residual connection~\cite{attention}:
\begin{equation}
Y = X + \mathrm{MultiHeadSelfAttn}(X) = X+\mathrm{Concat}\bigl[\mathrm{Head}_{1}\,\|\,\mathrm{Head}_{2}\,\|\;\cdots\;\|\mathrm{Head}_{h}\bigr]\,W^{O}\;\in\;\mathbb{R}^{N\times d}
\end{equation}
where $W^O \in \mathbb{R}^{d \times d}$ is a learnable on-site output projection matrix. Each attention head operates on a subspace of dimension $d_k = d/h$, where $h$ is the number of attention heads, and is computed via the standard scaled dot-product attention:
\begin{align}
    \mathrm{Head}_{i} \;=\mathrm{Attention}(Q_i,\,K_i,\,V_i)=\; \mathrm{softmax}\bigl(\frac{Q_{i}\,K_{i}^{\mathsf{T}}}{\sqrt{d_{k}}}\bigr)\;V_{i}
\;\in\;\mathbb{R}^{N\times d_{k}},
\end{align}
The query, key, and value matrices are obtained by independent linear projections of the input features:
\begin{align}
Q_{i} \;=\; X\,W^{Q}_{i}\;\in\;\mathbb{R}^{N\times d_{k}}, 
\quad
K_{i} \;=\; X\,W^{K}_{i}\;\in\;\mathbb{R}^{N\times d_{k}}, 
\quad
V_{i} \;=\; X\,W^{V}_{i}\;\in\;\mathbb{R}^{N\times d_{k}}.
\end{align}
where $W_i^Q, W_i^K, W_i^V \in \mathbb{R}^{d \times d_k}$ are head-specific learnable projection matrices for the $i$-th attention head. In this work, we adopt $h = 4$ attention heads with embedding dimension $d = 16$, resulting in $d_k = 4$ for each head.

The output $Y$ of the self-attention block is subsequently normalized along the feature dimension using layer normalization. The normalized features are then flattened into a one-dimensional vector $y$ and passed through an on-site two-layer multilayer perceptron (MLP) to generate the neuralized on-site tensors:
\begin{equation}
T_{\mathrm{NN}}^{[i]}(n) = \mathrm{LeakyRelu}(y W_1^{[i]}+ b_1^{[i]}) W_2^{[i]} + b_2^{[i]},
\end{equation}
where $W_1^{[i]} \in \mathbb{R}^{N d \times W}$, $W_2^{[i]} \in \mathbb{R}^{W \times \dim(T^{[i]})}$, $b_1^{[i]} \in \mathbb{R}^{W}$, and $b_2^{[i]} \in \mathbb{R}^{\dim(T^{[i]})}$ are learnable projection matrices and biases for the MLP associated with site $i$, and $W$ denotes the number of hidden neurons in the MLP.

The tensors in the fTNS are then modified as:
\begin{equation}
    T^{[i]}(n_i) \to T^{[i]}(n_i) + T_\text{NN}^{[i]}(n)
\end{equation}
which is similar to a backflow transformation~\cite{mpsbackflow,chen2024antn}, resulting in the configuration-dependent fTNS $|\Psi(n)\rangle$. The amplitude corresponding to the configuration $|n\rangle$ is then obtained by contracting the single layer fTN $\langle n|\Psi(n)\rangle$. 

{\it Model parameters initialization} ~In VMC optimization for the NN-fTNS Ansatz, the fTNS tensors $T^{[i]}(n_i)$ are initialized using the simple update (SU) method~\cite{Jiang2008}, while the neural network weights in the MLP blocks ($W_1^{[i]}$ and $W_2^{[i]}$) are initialized from a normal distribution $\mathcal{N}(0, 0.005)$, ensuring that the neural network components act as small perturbations to the underlying fTNS at the beginning of the optimization.

{\it Improving the neural network structure} ~In the final on-site MLP following the self-attention block, the second dimension of $W_1^{[i]}$ (equivalently, the first dimension of $W_2^{[i]}$) corresponds to the number of hidden neurons in the last layer of the neural network block interfacing with the tensor network elements, denoted as $W$, as illustrated in Fig.~\ref{fig:SM_model_arch}(c). The value of $W$ controls the bandwidth of information flow from the neural network to the tensor network. In the NN-fTNS models shown in Fig.~2 and Fig.~3 of the main text, $W$ is chosen to be proportional to the fTNS bond dimension $D$. As discussed and demonstrated in Fig.~4 of the main text, increasing $W$ can lead to a modest improvement in the performance of NN-fTNS, at the cost of enlarging the sizes of the weight matrices $W_1^{[i]}$ and $W_2^{[i]}$ with scaling $\dim(y)\times W$ and $W\times \dim(T^{[i]})$, respectively.

\subsection{Other NN based models}

We summarize below the architectural details of the various NN-based models used in Fig.~\ref{fig:model_comparison} of the main text.

For the pure NN model, we replace the fTNS structure in Fig.~\ref{fig:SM_model_arch}(a) with a two-layer MLP employing a shifted hyperbolic sine activation function, $f(x) = \sinh(x) + 1$~\cite{minSR}, which directly outputs a real scalar value $\mathrm{NN}(n)$ representing the amplitude for configuration $|n\rangle$. In VMC optimization, the pure NN model parameters are initialized randomly.

For the NN-SD model, we follow the standard neural network backflow construction~\cite{Luo2019, nnbfabinitio, nqshubbard}, replacing the fTNS in Fig.~\ref{fig:SM_model_arch} with a single-particle orbital matrix $M_{i,k} \in \mathbb{R}^{N_o \times N_e}$, where $N_o$ denotes the total number of spin orbitals and $N_e$ is the total number of electrons. Given a configuration $|n\rangle$, the NN modifies the orbital matrix to produce a configuration-dependent matrix $M(n)$. The rows of $M(n)$ corresponding to the occupied orbitals in $|n\rangle$ are selected to form a square matrix, whose determinant $\det(M_{i_o,j})$ yields the final amplitude, where $i_o$ labels the occupied orbitals. In VMC optimization, the NN-SD is initialized by first fully optimizing the Slater determinant orbital matrix (via minimizing its mean-field energy in the FH model), after which the NN parameters are initialized such that the neural network introduces only small perturbations to the fully optimized Slater determinant. The specific NN structures used are (1) NN-SD (MLP): a two-layer multilayer perceptron; (2) NN-SD (ViT): a Vision Transformer~\cite{vit} as shown in Fig.~\ref{fig:SM_model_arch}(a); (3) a convolutional neural network~\cite{HFPS,lecun2015deep}.

For the fPEPS$\times$NN-Jastrow model, given a configuration $|n\rangle$ and an fPEPS state $|\Psi\rangle$, we apply the pure NN model to compute a scalar factor $\mathrm{NN}(n)$, which is multiplied with the fPEPS amplitude to obtain the final amplitude $\mathrm{NN}(n) \times \langle n | \Psi \rangle$. In VMC optimization, the fTNS tensors are initialized using SU, and the NN model parameters are initialized randomly.

\subsection{VMC optimization curves for various Ans\"atze}
We show below an example of our VMC optimization procedure for various Ans\"atze on a small $4\times 2$ FH model. We can see the NN-fPEPS model requires fewer VMC optimization steps (around 500 steps) to converge than the NN-SD model (at least 1000 steps).
\begin{figure}[H]
    \centering
    \includegraphics[width=0.48\textwidth]{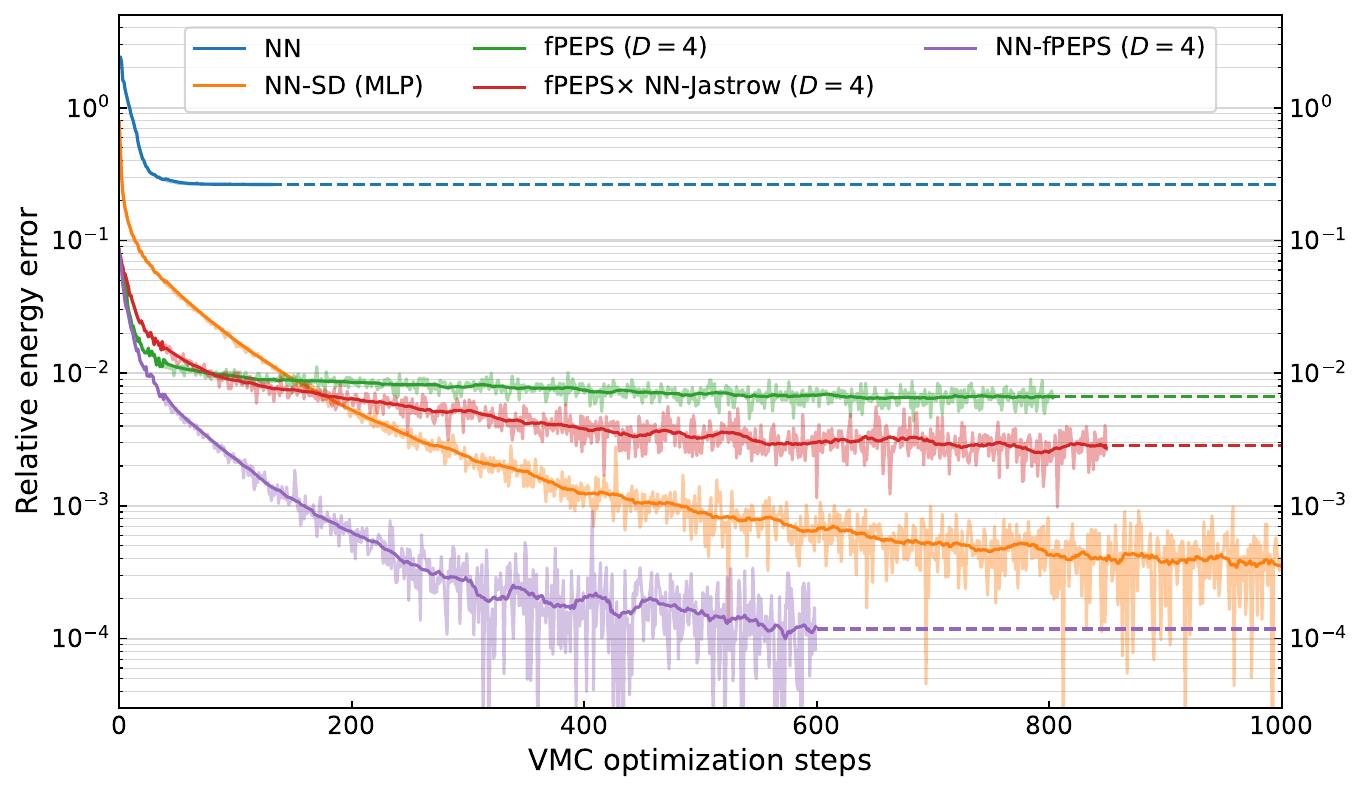}
    \caption{VMC optimization curve of various relevant Ans\"atze for the FH model at hole doping $n_h=1/4$ on a small $4\times 2$ square lattice. Errors are calculated relative to the exact ground state energy. NN: pure neural network parametrization; NN-SD: optimized Slater determinant with NN backflow (specific NN structure indicated in parentheses); fPEPS: pure fPEPS state; NN-fPEPS: neuralized fPEPS; fPEPS$\times$NN-Jastrow: Multiplicative NN-Jastrow factor on fPEPS.}
    \label{fig:vmc_optimization_compare}
\end{figure}

\section{S-2. Tensor network variational monte carlo}
\subsection{a. Variational Monte Carlo}
Here we provide a brief review of the variational Monte Carlo algorithm employed in this work, following the presentation in Refs.~\cite{tnfunc, Liu2021, Liu2017}.

In VMC, operator expectation values are estimated via importance sampling of configurations $|n\rangle = |n_1, \dots, n_N\rangle$, where $N$ is the system size. The energy of a Hamiltonian $H$ is evaluated as
\begin{equation}
    E=\frac{\langle\Psi|H|\Psi\rangle}{\langle\Psi|\Psi\rangle}=\frac{1}{\langle\Psi|\Psi\rangle}\sum_n |\langle n|\Psi\rangle|^2 \frac{\langle n|H|\Psi\rangle}{\langle n|\Psi\rangle} = \bigl\langle E_{loc}(n)\bigr\rangle_{n\sim p(n)}
    \label{local energy}
\end{equation}
where $p(n) = |\langle n | \Psi \rangle|^2 / \langle \Psi | \Psi \rangle$ and the local energy is defined as
\begin{equation}
    E_{loc}(n)=\frac{\langle n|H|\Psi\rangle}{\langle n|\Psi\rangle} = \sum_{n'}\langle n|H|n'\rangle\,\frac{\langle n'|\Psi\rangle}{\langle n|\Psi\rangle}.
\end{equation} 
We note that for a local Hamiltonian $H$ containing $O(N)$ local terms — for example, the Fermi-Hubbard model considered in this work — each configuration $|n\rangle$ is connected to $O(N)$ configurations $|n'\rangle$ with non-zero matrix elements $\langle n | H | n' \rangle$. Due to the locality of the Fermi-Hubbard model, $|n\rangle$ and $|n'\rangle$ differ at most on two neighboring sites. The energy is then estimated stochastically as the sample mean of $E_{\mathrm{loc}}(n)$ over configurations $n$ drawn from $p(n)$ using standard Markov chain sampling with the Metropolis algorithm. To generate new configurations, we employ the sequential nearest-neighbor spin-pair exchange algorithm~\cite{Liu2021}, which preserves the total spin $S_z$ quantum number during the sampling.

The energy gradient can similarly be evaluated using Monte Carlo sampling:
\begin{align}
    g =\frac{\partial E}{\partial \theta} &= \frac{1}{\langle \Psi|\Psi\rangle}\sum_n |\langle n|\Psi\rangle|^2\,\bigl( \frac{\langle\partial_\theta\Psi|n\rangle}{\langle \Psi|n\rangle}\bigr)\cdot E_{loc}(n) - \frac{\langle\Psi|H|\Psi\rangle}{\langle\Psi|\Psi\rangle}\cdot\frac{1}{\langle\Psi|\Psi\rangle}\sum_n |\langle n|\Psi\rangle|^2\,\frac{\langle\partial_\theta\Psi|n\rangle}{\langle \Psi|n\rangle} \\
    &= \bigl\langle E_{loc}(n) \cdot v_\theta(n)\bigr\rangle_{n\sim p(n)} - \bigl\langle E_{loc}(n)\bigr\rangle_{n\sim p(n)}\cdot\bigl\langle v_\theta(n)\bigr\rangle_{n\sim p(n)}
    \label{gradient}
\end{align}
where $v_\theta(n) = \langle \partial_\theta \Psi | n \rangle / \langle \Psi | n \rangle$ denotes the logarithmic derivative of the amplitude with respect to the model parameters $\theta$.

The variational state $|\Psi\rangle$ is optimized toward the ground state by gradient descent using stochastic reconfiguration (SR) optimization~\cite{SR1,SR2,SR3}. In SR, the model parameter update direction is preconditioned by the quantum geometric tensor (QGT)~\cite{QGT}, projecting the energy gradient along the direction of imaginary time evolution on the variational manifold. The parameter change $\mathrm{d}\theta$ is obtained by solving the linear system:
\begin{align}
R\,\frac{\mathrm{d}\theta}{\mathrm{d}\tau} = (S+\eta\,I)\,\frac{\mathrm{d}\theta}{\mathrm{d}\tau} = -g,
\end{align}
where $R = S + \eta I$, $\mathrm{d}\tau$ is a small positive time step, $S$ is the QGT, and $\eta I$ is a small diagonal shift introduced to stabilize the linear solver. Solving the equation yields
\begin{equation}
\mathrm{d}\theta = - (R^{-1} g)\, \mathrm{d}\tau,
\end{equation}
where $R^{-1} g$ is the SR preconditioned gradient along the imaginary time evolution direction on the variational manifold. The QGT $S$ is estimated via Monte Carlo sampling as
\begin{equation}
    S_{ij} = \bigl\langle v_{\theta_i}^*(n)\,v_{\theta_j}(n)\bigr\rangle_{n\sim p(n)} - \bigl\langle v_{\theta_i}^*(n)\bigr\rangle_{n\sim p(n)} \bigl\langle v_{\theta_j}(n)\bigr\rangle_{n\sim p(n)}
\end{equation}
SR can be viewed as a stochastic realization of the imaginary time-dependent variational principle (TDVP)~\cite{tdvp}. Finally, model parameters are updated via gradient descent using the preconditioned gradient, $\theta \to \theta - \gamma R^{-1} g$, where $\gamma$ denotes the learning rate. In this work, we adopt a dynamic step size schedule, decaying $\gamma$ from $0.05$ to $0.01$ over the course of the VMC training.

In all of our VMC calculations, the neural networks and the fermionic bias (either Slater determinant or fTNS) are optimized jointly during the variational optimization.

\subsection{b. Partial tensor network contraction reuse}

\begin{figure}[htbp]
    \centering
    \includegraphics[width=0.8\textwidth]{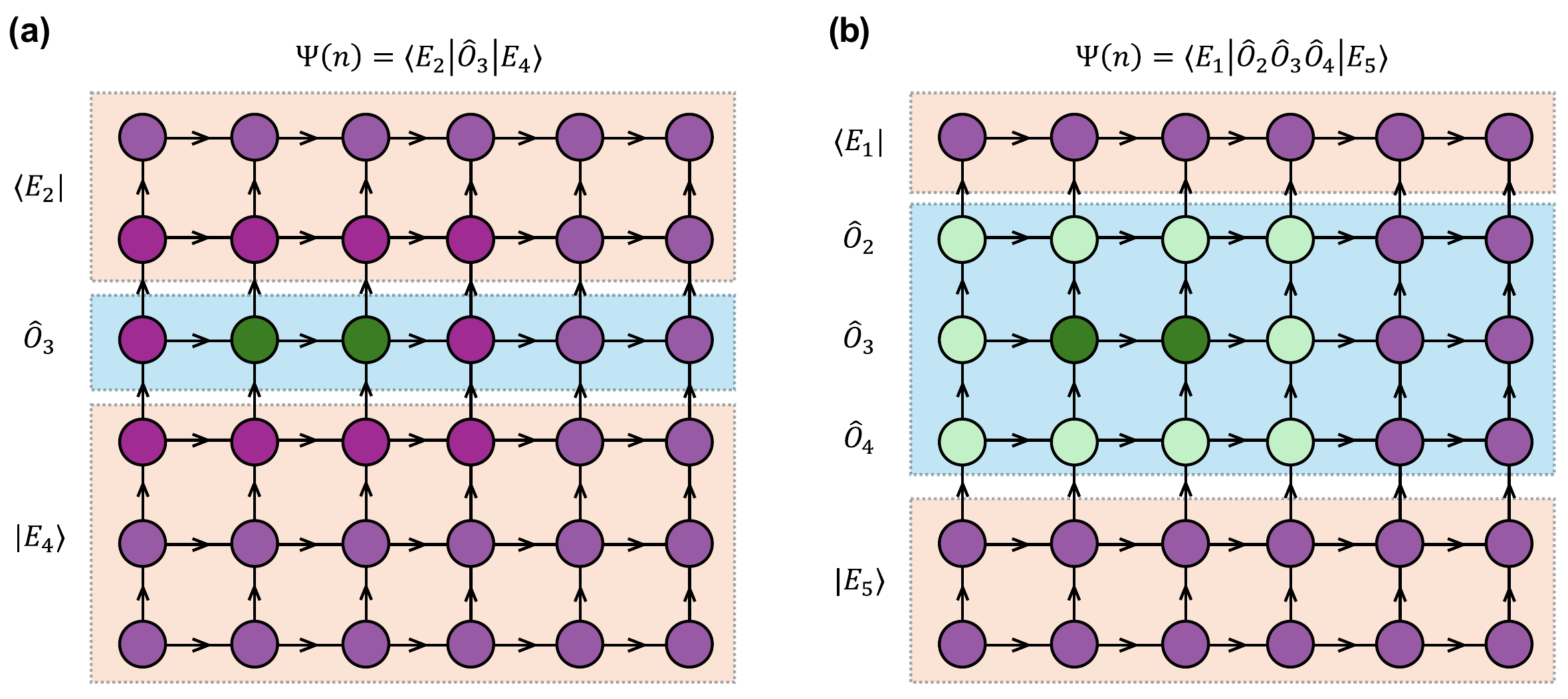} 
    \caption{Tensor network contraction reuse for (a) fPEPS and (b) NN-fPEPS with finite neural network range $R=1$. Configuration updates occur on the dark green sites, while the light green sites indicate tensors affected by configuration changes on the dark green sites. The intermediates $|E_i\rangle$ are computed using the standard boundary-MPS method~\cite{Verstraete2008, Verstraete2004}.}
    \label{fig:SM_TN_reuse}
\end{figure}

In Markov Chain VMC calculations, a sequence of configurations that differ locally is typically generated, for example, during the evaluation of local energies for a local Hamiltonian. In the boundary-MPS contraction of TN amplitudes corresponding to these configurations, a significant portion of the computation—associated with partial TN contractions—can be reused, as illustrated in Fig.~\ref{fig:SM_TN_reuse}. 

As shown in Fig.~\ref{fig:SM_TN_reuse}(a), for a pure fTNS, each local tensor depends only on its on-site configuration. Consequently, modifying the configurations in the third row does not affect the environment tensor networks $\langle E_2|$ and $|E_4\rangle$, which can thus be reused for contracting amplitudes $\langle E_2|\hat{O}_3|E_4\rangle$ that differ solely in the third-row tensors.

In contrast, as shown in Fig.~\ref{fig:SM_TN_reuse}(b), for a NN-fPEPS with finite neural network range $R$ (with $R=1$ illustrated), each local tensor depends not only on its own configuration but also on those of neighboring sites within range $R$. Consequently, configuration changes in the third row also affect tensors in the second and fourth rows. The environment tensor networks $\langle E_1|$ and $|E_5\rangle$ remain unaffected and can be reused for contracting amplitudes $\langle E_1|\hat{O}_2\hat{O}_3\hat{O}_4|E_5\rangle$, where only the tensors in rows 2 to 4 are varied.

\section{S-3. Quantum state fitting}
As a supplement to the main text and to further demonstrate the enhanced expressivity of NN-fPEPS, we perform supervised wavefunction optimization (SWO)~\cite{SWO1, Zejun2024} for quantum state learning. In this approach, we stochastically fit an NN-fPEPS with bond dimension $D=4$ to a pure fPEPS with a larger bond dimension $D=8$, by minimizing the negative logarithmic fidelity loss:
\begin{equation}
    \mathcal{L} = -\log F = -\log \frac{\langle\Psi_\theta|\Phi_t\rangle\langle\Phi_t|\Psi_\theta\rangle}{\langle\Psi_\theta|\Psi_\theta\rangle\langle\Phi_t|\Phi_t\rangle}
\end{equation}
where $|\Psi_\theta\rangle$ denotes the NN-fPEPS and $|\Phi_t\rangle$ represents the target pure fPEPS. For comparison, we also perform fitting using a pure fPEPS of the same bond dimension $D=4$.

\begin{figure}[htbp]
\centering
\includegraphics[width=0.45\textwidth]{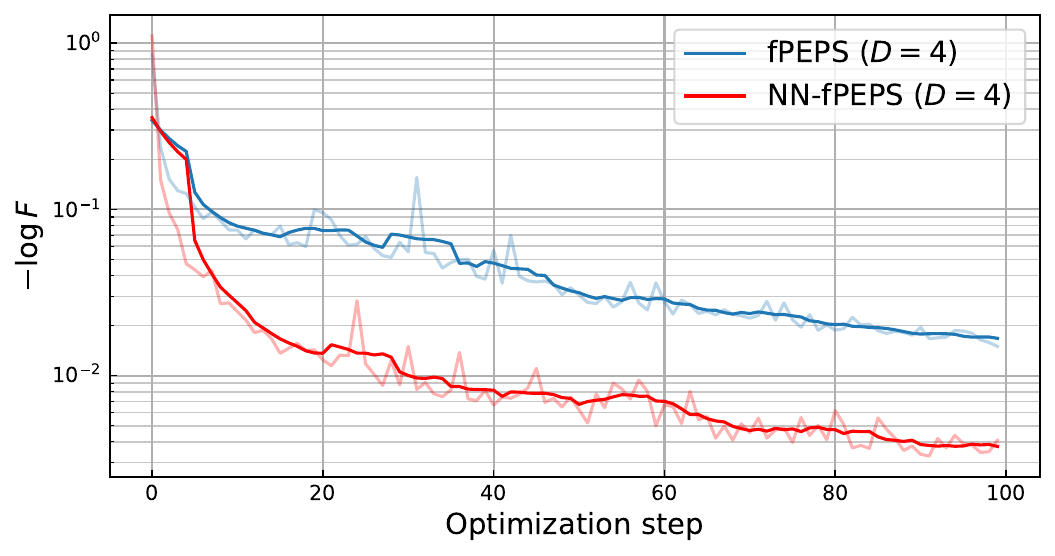}
\caption{SWO state fitting loss curve for the ground state of the 2D Fermi-Hubbard model on a $4\times 2$ square lattice at $n_h=1/8$ hole doping. The target state is an optimized fPEPS with bond dimension $D=8$ (relative energy error $\epsilon = 10^{-4}$). In each optimization step, $10,000$ samples are drawn from $|\Phi_t\rangle$ to estimate the fidelity, which are used as the training set for minimizing the negative logarithmic fidelity $\mathcal{L} = -\log F$.}
\label{fig:state_fitting}
\end{figure}

We observe a clear separation in the quantum state fitting loss curves: NN-fPEPS rapidly achieves higher fidelity compared to the pure fPEPS, explicitly demonstrating its superior expressivity; see Fig.~\ref{fig:state_fitting}.

\section{S-4. Computational cost analysis}

\begin{figure}[H]
\centering
\includegraphics[width=0.6\textwidth]{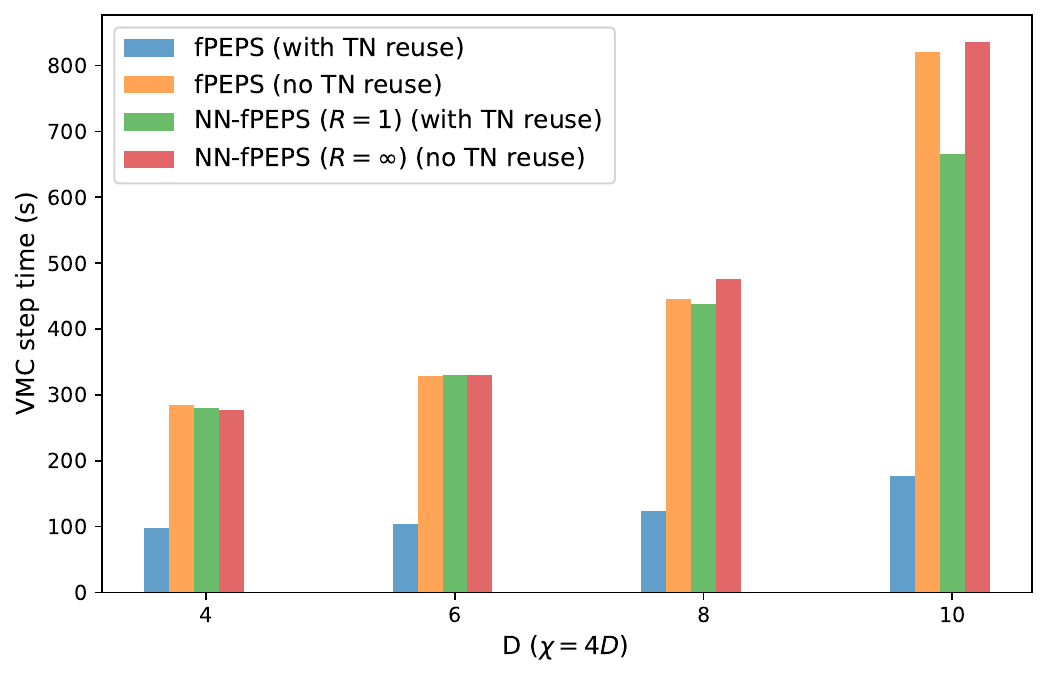}
\caption{Time cost comparison for a single VMC step (20 samples per step) across different fPEPS bond dimensions $D$ on a $6 \times 6$ Fermi-Hubbard model. Each timing is averaged over 5 independent trials on a single CPU core. The boundary-MPS contraction bond dimension is chosen as $\chi = 4D$. We compare pure fPEPS with and without TN reuse, NN-fPEPS with finite neural network range ($R=1$) with reuse, and NN-fPEPS with global range ($R=\infty$) without reuse.}
\label{fig:SM_VMC_cost_comparison}
\end{figure}

For the pure fTNS model and NN-fTNS model without TN contraction reuse, the computational cost of each VMC optimization step scales quadratically as $O(N^2)$ with system size $N$. The contraction of a single-layer amplitude $\langle n | \Psi \rangle$ has cost $O(N)$, which is the dominant cost in the computation of NN-fTNS amplitudes. Each VMC iteration mainly consists of two parts: sampling new configurations and computing local energies. In the sequential exchange sampling algorithm adopted in this work~\cite{Liu2021}, one sequential sweep over the system requires $O(N)$ local configuration updates to generate a new configuration. Each local update, without TN contraction reuse, has cost $O(N)$, leading to an overall sampling cost of $O(N^2)$ per sweep. For local Hamiltonians such as the Fermi-Hubbard model, computing the local energy for a given configuration $|n\rangle$ requires evaluating amplitudes for $O(N)$ connected configurations $|n_c\rangle$, where $\langle n_c | H | n \rangle \neq 0$, again leading to $O(N^2)$ cost when TN contraction reuse is not employed. Including the cost of boundary-MPS contraction for a single-layer amplitude TN, which scales as $O(D^4 \chi^2)$ (with $D$ the fTNS bond dimension and $\chi$ the auxiliary bond dimension), and adopting the typical choice $\chi = O(D)$, the total leading cost per VMC step becomes $O(M D^6 N^2)$~\cite{Liu2025}, where $M$ denotes the Monte Carlo sample size. 

In practice, the VMC sampling can be efficiently parallelized by distributing multiple Markov chains across CPU cores in an MPI program, reducing the wall-clock time to $O(M D^6 N^2 / N_{\mathrm{chain}})$, where $N_{\mathrm{chain}}$ is the number of parallel chains. In this work, we typically use $M$ of a few tens of thousands and $N_{\mathrm{chain}} \approx 500$.

For fTNS based models where local tensors only have local configuration dependence, partial TN contraction can be reused for computation of related amplitudes, as illustrated in Fig.\ref{fig:SM_TN_reuse}. For pure fTNS, this reuse strategy allows the VMC cost to be reduced from $O(N^2)$ to nearly $O(N)$ in principle. For example, in the computation of the local energy for a configuration $|n\rangle$ using boundary MPS contraction, one may first cache all partial TN contraction intermediates (environment MPS $|E_i\rangle$ in Fig.\ref{fig:SM_TN_reuse}) for $\langle n|\Psi\rangle$. Then, to compute $\langle n_c|\Psi\rangle$ for a locally connected configuration $|n_c\rangle$, only a three-row TN $\langle E_{i-1}|\hat{O}_i|E_{i+1}\rangle$ needs to be contracted, which scales as $O(\sqrt{N})$ for a square lattice. If additional TN reuse along the column direction is employed, each $|n_c\rangle$ requires contracting only a small plaquette TN with constant cost $C_{n_c}$ independent of system size, reducing the total cost for local energy evaluation to $O(C_{n_c} N)$~\cite{Liu2017, Liu2021, tnfunc}. For NN-fTNS models with finite neural network range $R$, where each local tensor depends on neighboring configurations within range $R$, a similar partial TN contraction reuse scheme can be applied. A comparison of the VMC cost scaling for the above mentioned models as well as conventional fermionic NQS is listed in Table.~\ref{tab:cost_comparison_simple}.

As illustrated in Fig.~\ref{fig:SM_VMC_cost_comparison} for a $6 \times 6$ square lattice, TN reuse (applied along the row direction) significantly reduces the VMC step cost for pure fPEPS across various bond dimensions, while yielding a more modest cost reduction for NN-fPEPS with neural network range $R=1$ at larger bond dimensions. We expect the benefit of TN reuse for NN-fPEPS ($R=1$) to become more pronounced as the system size increases. As discussed in the main text, we emphasize that even employing a minimal finite-range neural network with $R=1$ yields ground-state energy accuracies that are substantially better than those achieved by pure fTNS.

\begin{table}[H]
    \centering % Center the table on the page
    \caption{Comparison of VMC cost scaling with respect to system size $N$ for various fermionic Ans\"atze.}
    \label{tab:cost_comparison_simple}
    \begin{tabular}{llcc}
        \toprule % Top rule of the table
        \textbf{Model} & \textbf{VMC cost scaling} \\
        \midrule % Middle rule, separating header from data
        NN-fTNS (no reuse)     & $\mathcal{O}(N^2)$ \\
        NN-fTNS (with reuse)     & $\geq O(N),\ <\mathcal{O}(N^2)$ \\
        fTNS (with reuse)       & $\mathcal{O}(N)$ \\
        NN-SD (Ref.~\cite{Luo2019})  & $\mathcal{O}(N^4)$ \\
        HFPS (Ref.~\cite{HFPS})  & $\mathcal{O}(N^3)$ \\
        \bottomrule % Bottom rule of the table
    \end{tabular}
    \begin{flushleft} % Align notes to the left below the table
    % \textsuperscript{a}{Data extracted from published results in reference [XX].}\\
    % \textsuperscript{b}{Density Matrix Renormalization Group result, often considered a benchmark for 1D-like systems.}\\
    % \textsuperscript{c}{Scaling is exponential in the smaller dimension of the 2D lattice.}
    \end{flushleft}
\end{table}

\section{S-5. Energy tables}

Below we present the raw energy values used to compute the relative errors reported in the main text. The pure NN model and the NN backflow model employed in this work have total model sizes comparable to that of the smallest NN-fTNS with bond dimension $D=4$.
\subsection{a. One-dimensional Fermi-Hubbard model}

\textit{$L=24$ at $n_h=1/6$ hole doping ($N_{\uparrow}=N_{\downarrow}=10$)}

\begin{table}[H]
\centering
\resizebox{0.8\textwidth}{!}{
\renewcommand{\arraystretch}{1.5}
\begin{tabular}{|c|c|c|c|}
\hline

\diagbox{Model}{D} & $4$ &  $6$ & $8$  \\ \hline
fMPS-SU  & $-0.490(2)$  & $-0.531(7)$ & $-0.532(8)$  \\ \hline
fMPS-GO  & $-0.5037(8)$  & $-0.5348(0)$ & $-0.5470(4)$  \\ \hline
NN-fMPS-GO  & $\mathbf{-0.56143\pm 6\times 10^{-5}}$ & $\mathbf{-0.56212\pm 4\times 10^{-5}}$ & $\mathbf{-0.56244\pm 3\times 10^{-5}}$  \\ \hline

\end{tabular}
}
\caption{Energy per site of pure fMPS and NN-fMPS for the Fermi-Hubbard model on an $L=24$ one-dimensional chain at hole doping $n_h=1/6$ ($N_e=20$, $N_{\uparrow}=N_{\downarrow}=10$). The DMRG reference energy per site is $-0.56309964$, obtained using \texttt{pyBlock2}~\cite{block2} with fMPS bond dimension 1000 and truncated weight $\mathrm{d}w = 1.6 \times 10^{-10}$.}
\label{tab:L=24_energy}
\end{table}

\subsection{b. Two-dimensional Fermi-Hubbard model}

\textit{$L_x=L_y=4$ at half-filling ($N_{\uparrow}=N_{\downarrow}=8$)}
\begin{table}[H]
\centering
\resizebox{0.8\textwidth}{!}{
\renewcommand{\arraystretch}{1.5}
\begin{tabular}{|c|c|c|c|c|}
\hline
\diagbox{Model}{D} & $4$ &  $6$ & $8$ & $14$ \\ \hline
fPEPS-SU~\cite{Liu2025}       & $-0.4174(2)$   & $-0.4204(6)$      & $-0.4238(5)$ & $-0.4242(5)$      \\ \hline
fPEPS-GO~\cite{Liu2025}       & $-0.42115(3)$   & $-0.42416(7)$      & $-0.42508(4)$ & $-0.425508(5)$      \\ \hline
NN-fPEPS-GO & $\mathbf{-0.42535\pm 4\times 10^{-5}}$   & $\mathbf{-0.42541\pm 3\times 10^{-5}}$      & $\mathbf{-0.42551\pm 2\times 10^{-5}}$      & - \\ \hline
\end{tabular}
}
\caption{Energy per site of pure fPEPS and NN-fPEPS for the Fermi-Hubbard model on a $4\times 4$ square lattice at half-filling ($N_e=16$, $N_{\uparrow}=N_{\downarrow}=8$). The DMRG reference energy per site is $-0.42552590$~\cite{Liu2025}. The pure NN model yields $E_{\mathrm{NN}} = -0.3290$, while the NN backflow baseline calculation gives $E_{\mathrm{NN-SD\ (MLP)}} = -0.4157$ and $E_{\mathrm{NN-SD\ (ViT)}} = -0.4247$.}
\label{tab:4x4_half_energy}
\end{table}

\textit{$L_x=L_y=4$ at $n_h=1/8$ hole doping ($N_{\uparrow}=N_{\downarrow}=7$)}
\begin{table}[H]
\centering
\resizebox{0.8\textwidth}{!}{
\renewcommand{\arraystretch}{1.5}
\begin{tabular}{|c|c|c|c|c|c|c|}
\hline
\diagbox{Model}{D} & $4$ & $6$ & $8$ & $10$ & $14$ \\ \hline
fPEPS-SU~\cite{Liu2025} & $-0.5497(6)$   & $-0.5850(4)$      & $-0.6016(3)$ & $-0.6104(4)$ &$-0.6154(4)$     \\ \hline
fPEPS-GO~\cite{Liu2025} & $-0.5892(2)$   & $-0.61225(7)$      & $-0.62314(6)$ & $-0.62725(4)$ &$-0.63206(1)$     \\ \hline
NN-fPEPS-GO & $\mathbf{-0.6247\pm 2\times 10^{-4}}$ & $\mathbf{-0.6301\pm 2\times 10^{-4}}$      & $\mathbf{-0.63215\pm 7\times 10^{-5}}$      & $\mathbf{-0.63231\pm 5\times 10^{-5}}$ & -\\ \hline
\end{tabular}
}
\caption{Energy per site of pure fPEPS and NN-fPEPS for the Fermi-Hubbard model on a $4\times 4$ square lattice at $1/8$ hole doping ($N_e=14$, $N_{\uparrow}=N_{\downarrow}=7$). The DMRG reference energy is $-0.63261830$~\cite{Liu2025}. The pure NN model yields $E_{\mathrm{NN}} = -0.5391$, while the NN backflow baseline calculation gives $E_{\mathrm{NN-SD\ (MLP)}} = -0.6125$ and $E_{\mathrm{NN-SD\ (CNN)}} = -0.6288$.}
\label{tab:4x4_doped_energy}
\end{table}

\textit{$L_x=L_y=6$, at $n_h=1/9$ hole-doping ($N_{\uparrow}=N_{\downarrow}=16$)}
\begin{table}[H]
\centering
\resizebox{0.8\textwidth}{!}{
\renewcommand{\arraystretch}{1.5}
\begin{tabular}{|c|c|c|c|c|}
\hline
\diagbox{Model}{D} & $4$ &  $6$ & $8$ & $16$ \\ \hline
fPEPS-SU\cite{Liu2025}  & $-0.5786(4)$  & $-0.6200(6)$      & $-0.6312(5)$      & $ -0.6470(4)$\\ \hline
fPEPS-GO\cite{Liu2025}  & $-0.6007(7)$  & $-0.6309(4)$      & $-0.6415(6)$      & $ -0.65327(5)$\\ \hline
NN-fPEPS-GO & $\mathbf{-0.6370\pm 2\times 10^{-4}}$ & $\mathbf{-0.6503\pm 1\times 10^{-4}}$      & $\mathbf{-0.6535\pm 1\times 10^{-4}}$      & -\\ \hline
\end{tabular}
}
\caption{Energy per site of pure fPEPS and NN-fPEPS for the Fermi-Hubbard model on a $6\times 6$ square lattice at $1/9$ hole doping ($N_e=32$, $N_{\uparrow}=N_{\downarrow}=16$). The DMRG reference energy per site is $-0.6573063$~\cite{Liu2025}.}
\label{tab:6x6_energy}
\end{table}

%TC:endignore
\end{document}